\documentclass[aip,apl,reprint ]{revtex4-1}

\usepackage{graphicx}
\usepackage{epstopdf}
\usepackage{dcolumn}
\usepackage{bm}
\usepackage{amsmath}
\usepackage{amssymb}
\usepackage{wasysym}
\usepackage{color}
\usepackage{amssymb} 
\usepackage{dsfont} 

\begin{document}

\title{Tuning the thermal conductivity of graphene nanoribbons by edge passivation and isotope engineering: a molecular dynamics study} 



\author{Jiuning Hu}
\email[]{hu49@purdue.edu}
\altaffiliation{School of Electrical and Computer Engineering, Birck
Nanotechnology Center, Purdue University, West Lafayette, Indiana
47907, USA}

\author{Stephen Schiffli}
\altaffiliation{School of Electrical and Computer Engineering, Purdue University, West Lafayette, Indiana
47907, USA}

\author{Ajit Vallabhaneni}
\altaffiliation{School of Mechanical Engineering, Purdue University, West Lafayette, Indiana
47907, USA}

\author{Xiulin Ruan}
\altaffiliation{School of Mechanical Engineering, Birck
Nanotechnology Center, Purdue University, West Lafayette, Indiana
47907, USA.}

\author{Yong P. Chen}
\email[]{yongchen@purdue.edu}
\altaffiliation{Department of Physics, School of Electrical and
Computer Engineering, Birck Nanotechnology Center, Purdue
University, West Lafayette, Indiana 47907, USA}



\date{\today}

\begin{abstract}
Using classical molecular dynamics simulation, we have studied the effect of edge-passivation by hydrogen (H-passivation) and isotope mixture (with random or supperlattice distributions) on the thermal
conductivity of rectangular graphene nanoribbons (GNRs) (of several nanometers in size). We find that the thermal
conductivity is considerably reduced by the edge H-passivation. We also
find that the isotope mixing can reduce the thermal
conductivities, with the supperlattice distribution giving rise to more reduction than the random distribution. These
results can be useful in nanoscale engineering of thermal transport and heat
management using GNRs.
\end{abstract}


\maketitle 


Graphene\cite{Geim07,Castro08} is a monolayer of graphite with a honeycomb
lattice structure. It exhibits many unique properties and has drawn
intense attention in the past few years. The unusual
electronic properties of graphene are promising in many fundamental studies
and applications, e.g., the ultrahigh electron mobility\cite{Castro08} and the tunable band gap
and magnetic properties by the size and edge chirality of GNRs.\cite{Nakada96, Son06, Han07, Chen07}
Graphene also has remarkable thermal properties. The measured value of thermal conductivity of graphene reaches as high as several thousand of W/m-K,\cite{Alexander08, Cai10, Faugeras10, Luis10} among the highest values of known materials. Previous studies\cite{Hu09,Jiang09,Xu09} show that the thermal transport in GNRs depends on the edge chirality of GNRs. In realistic graphene samples, the edges are often passivated\cite{Park09,Boukhvalov09,Lu09} and the isotope composition can be controlled.\cite{Li09b} Motivated by these, we study the effect of the edge H-passivation and various isotope distributions on the thermal transport in GNRs.
We find that the thermal conductivity can be reduced by the edge H-passivation and tuned
by the isotope distributions. Our study is useful in
nanoscale control and management of thermal transport by engineering
the chemical composition of GNRs.

\begin{figure}
  \includegraphics[width=0.42\textwidth]{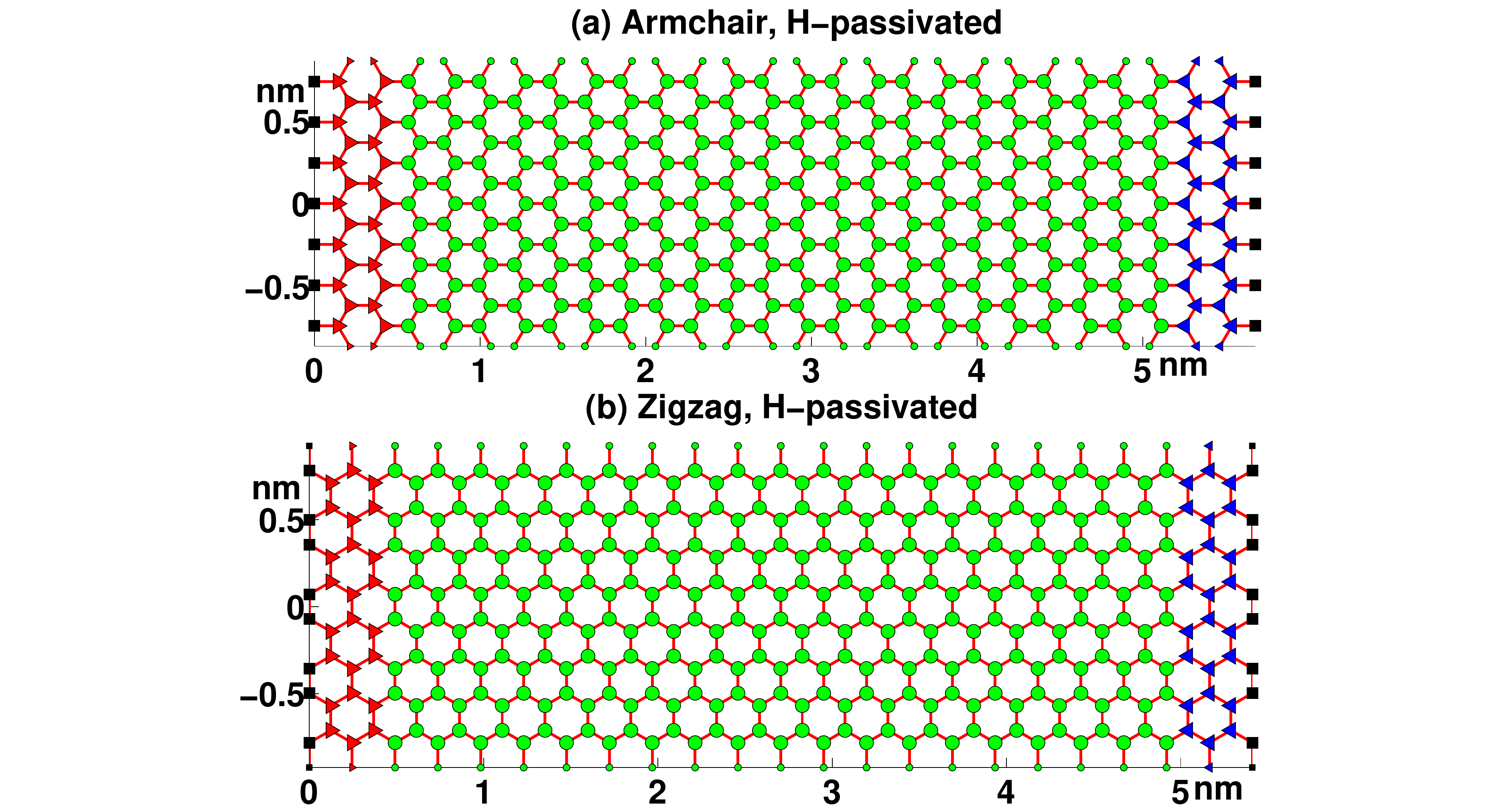}\\
  \caption{Structure of hydrogen-passivated armchair (a) and
zigzag (b) GNRs. The hydrogen atoms are denoted by smaller symbols while the $^{12}$C atoms are denoted by larger
ones. 
$\blacksquare$ denotes fixed boundary atoms. 
$\color{red}\blacktriangleright$ ($\color{blue}\blacktriangleleft$) denotes atoms in the left (right) thermostat.  
$\color{green}\CIRCLE$ denotes the remain atoms in the bulk.
}\label{fig1}
\end{figure}

In this work, we employ the classical molecular dynamics (MD, similar to the method in Ref.~\onlinecite{Hu09}) to calculate the thermal conductivities of GNRs.
We use the Brenner
potential,\cite{Brenner90} which incorporates the many-body carbon-carbon and carbon-hydrogen interactions by introducing a fractional number of covalent bonds. This method has been successfully applied to many carbon-based systems,\cite{Gong08,Zhang09} especially to graphene.\cite{Hu09,Wang09, Ong10} The structures of GNRs are shown in Fig.~\ref{fig1} (with edge H-passivation) and the insets of Fig.~\ref{fig3} (with isotope mixtures). We
use fixed boundary conditions, i.e., the atoms denoted by 
squares are fixed at their equilibrium positions. The atoms denoted by left- and
right-pointing triangles are placed in two Nos\'e-Hoover\cite{Nose84,Hoover85} thermostats at different temperatures. The thermal
conductivity can be calculated from the Fourier law $\kappa=Jd/(\Delta Twh)$, where $\Delta T$ is the temperature
difference (chosen to be in the linear response regime\cite{Hu09b}) between two thermostats, $J$ is the resulting thermal current, $d$ is the length, $w$ is the width and $h$(=0.335 nm) is the thickness of GNRs, respectively. Calculations presented below are performed for representative GNRs with $d\sim$ 6nm and $w\sim$ 1.5 nm. All the calculated thermal conductivities are normalized by $\kappa_0$. $\kappa_0$ is the thermal conductivity calculated at 100 K for the pure $^{12}$C GNR with armchair edge and without H-passivation (shown in Figure (a) in Ref.~\onlinecite{supp}). Although the specific value of $\kappa_0$ depends on the GNR size,\cite{Hu09b} the choice of thermostat and boundary condition in MD simulation,\cite{Chen10,Jiang09} we have checked that our conclusions and the qualitative behavior of $\kappa$ discussed below do not. The
equations of motion for atoms labeled by triangles in either left or right Nos\'e-Hoover thermostat are:
\begin{equation}
\frac{d}{dt}\mathbf p_i=\mathbf F_i-\gamma\mathbf p_i,
\frac{d}{dt}\gamma=\frac{T(t)-T_0}{\tau^2T_0}
\end{equation}
where $i$ runs over all the atoms in the thermostat,  $\mathbf p_i$ is the momentum
of the $i$-th atom, $\mathbf F_i$   is the total force acting on the $i$-th atom, $\gamma$
is the dynamic parameter of the thermostat with initial value of
zero,  $\tau$  is the relaxation time of the thermostat, $T(t)\equiv\frac{2}{3Nk_B}\sum_i\frac{\mathbf p_i^2}{2m_i}$ is the actual
temperature of atoms in the thermostat at time instant $t$ ,  $T_0=T_L$ (or $T_R$ ) is
the desired temperature of the left (or right) thermostat, $N$  is the
number of atoms in the thermostat, $k_B$  is the Boltzmann constant and $m_i$
is the mass of the $i$-th atom. The atoms labeled by circles obey the Newton's law of motion with $\gamma=0$. The carbon-carbon potential is the same for all carbon isotopes.

\begin{figure}
  \includegraphics[width=0.45\textwidth]{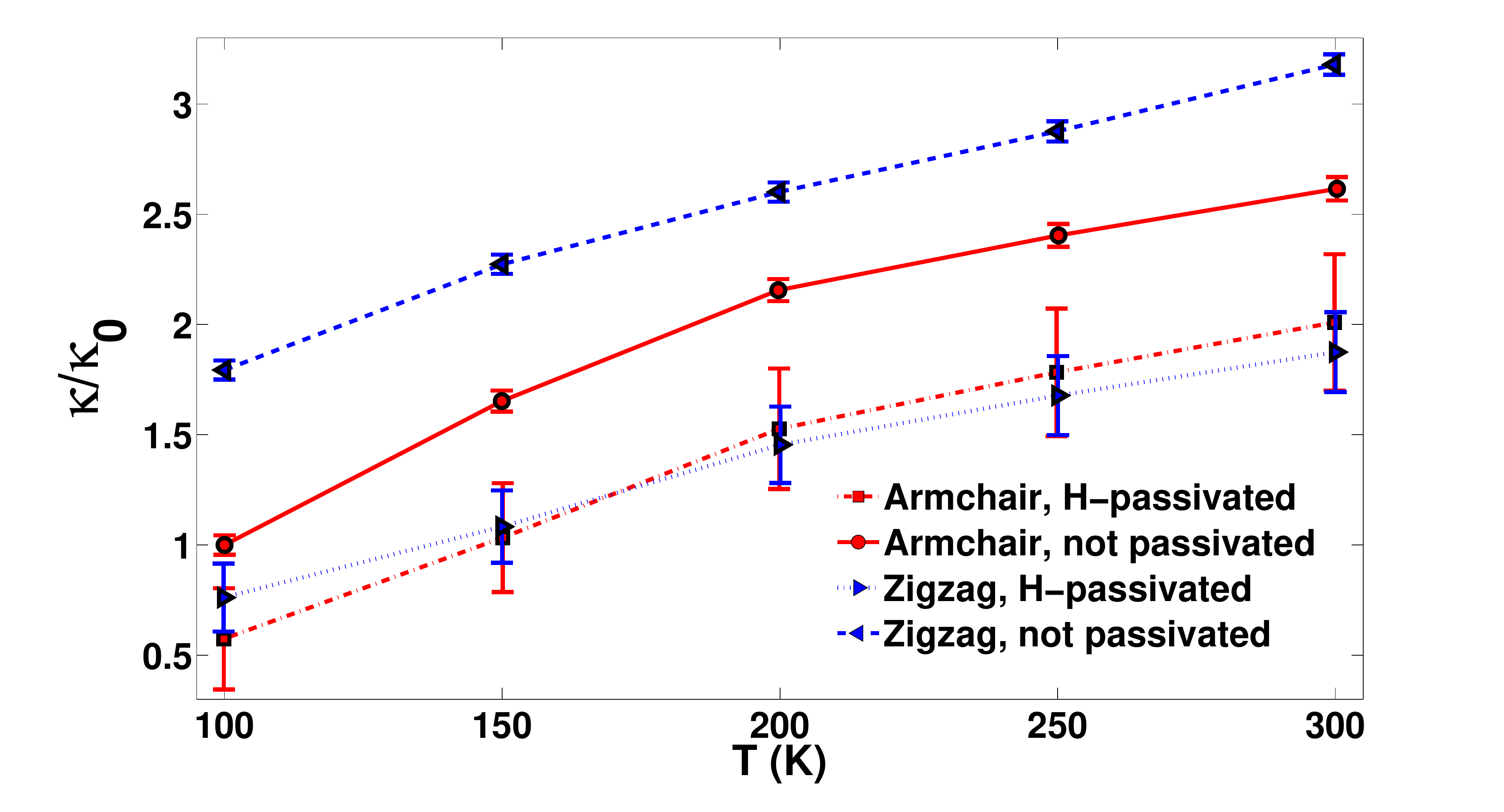}\\
  \caption{Temperature dependent thermal conductivity of
GNRs with and without edge H-passivation.}\label{fig2}
\end{figure}

\begin{figure}
  \includegraphics[width=0.48\textwidth]{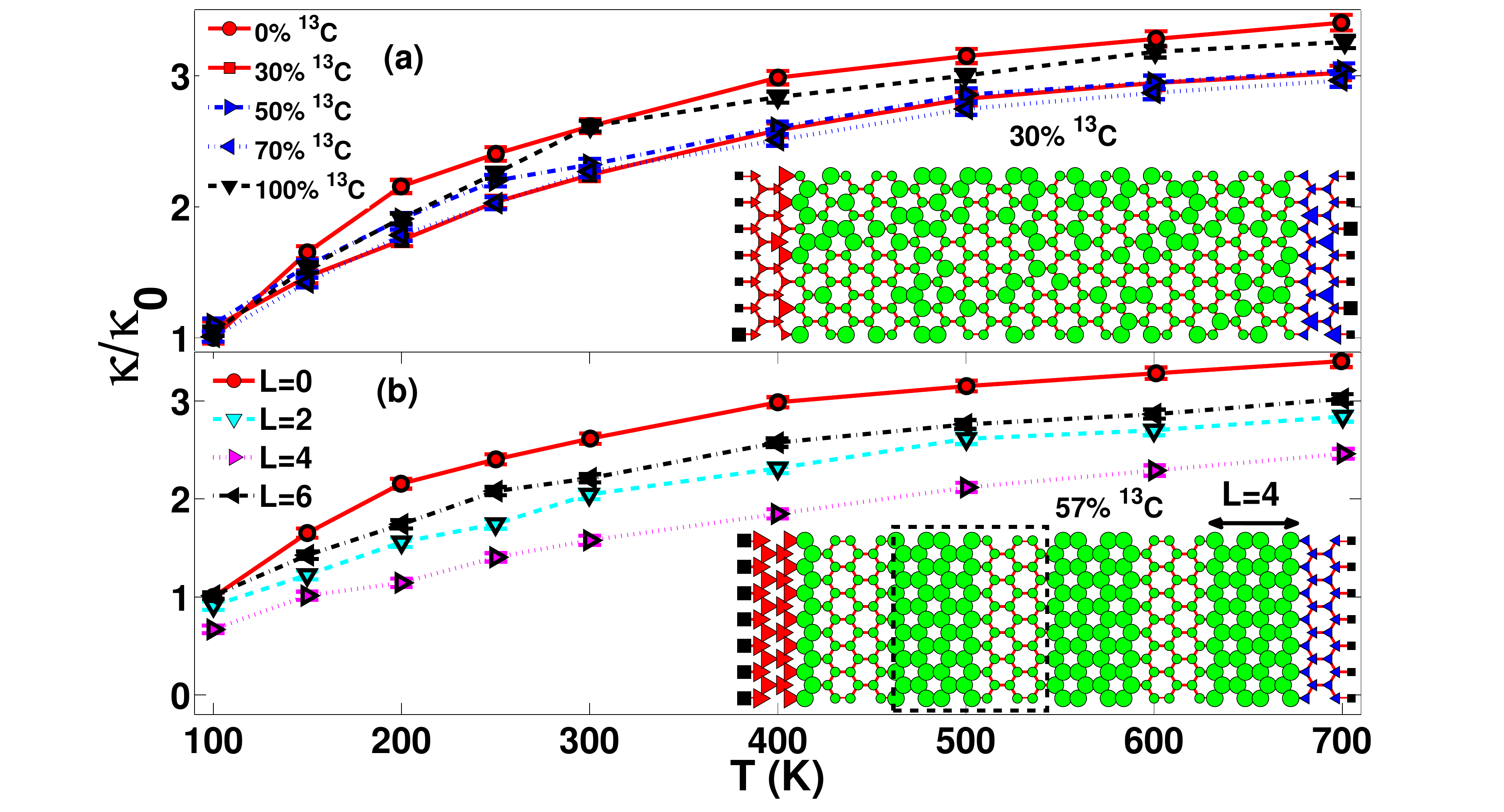}\\
  \caption{Temperature dependent thermal conductivity of GNRs
with $^{13}$C isotopes distributed (a) randomly and (b) in a
supperlattice structure. The insets show the corresponding typical
structures of GNRs with the same meaning of symbols as that in Fig.~\ref{fig1}. The larger (smaller) symbols denote $^{13}$C ($^{12}$C) atoms.}\label{fig3}
\end{figure}

First, we study the temperature dependent thermal
conductivity of H-passivated GNRs. Fig.~\ref{fig1} shows the armchair and
zigzag GNRs with top and bottom edges H-passivated. As shown in Fig.~\ref{fig2}, the edge H-passivation significantly reduces
the thermal conductivity, compared to the non-passivated GNRs. A recent study\cite{Evans10} using equilibrium MD has obtained similar conclusions. We note that the error bars (related to molecular dynamics fluctuations) for
H-passivated GNRs are considerably larger than that for the non-passivated GNRs, probably due to the much smaller mass of
hydrogen atoms. 

\begin{figure}
  \includegraphics[width=0.45\textwidth]{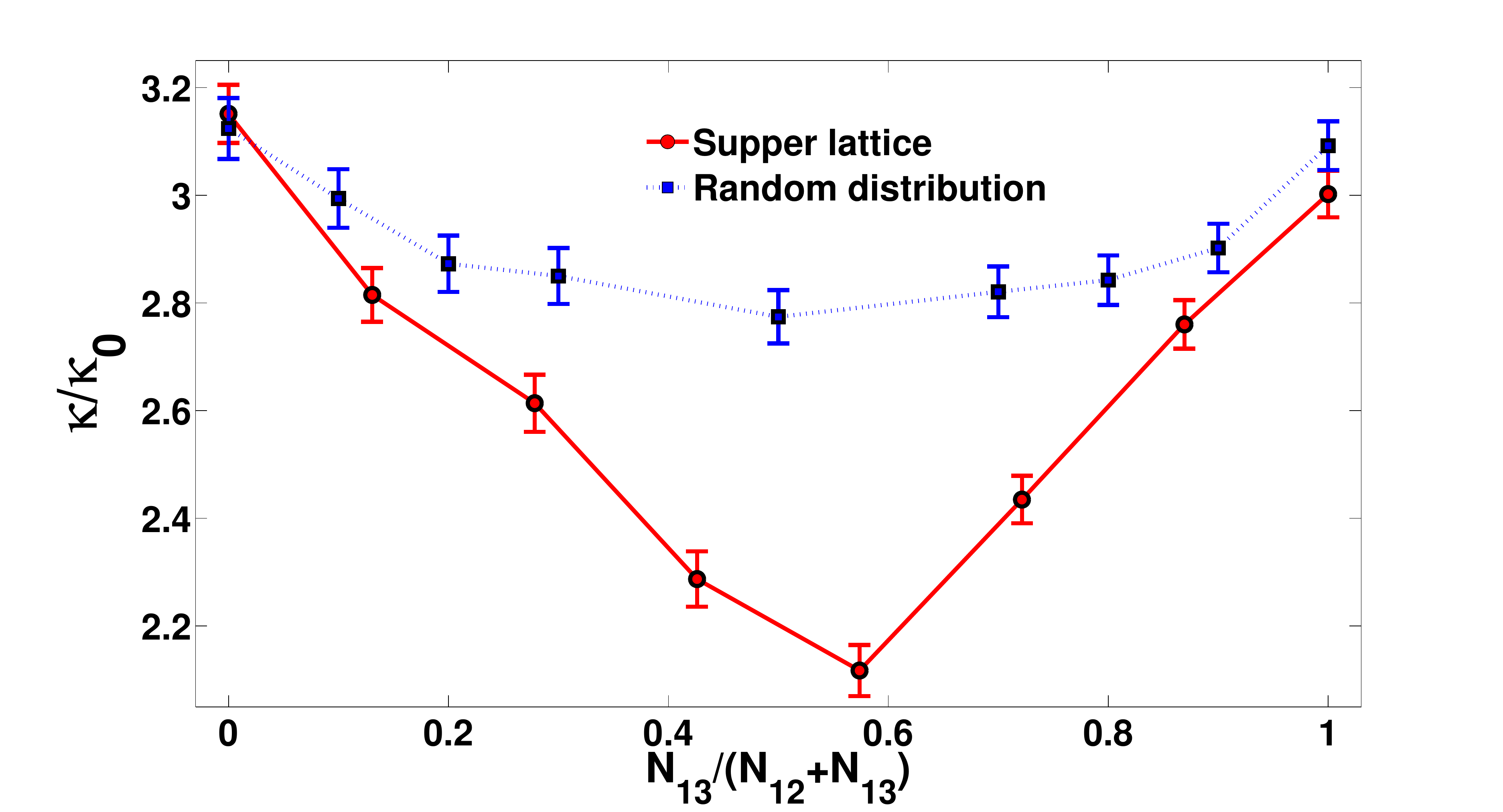}\\
  \caption{Thermal conductivity as a function of the $^{13}$C
concentration for supperlattice (solid line) and random (dashed line) isotope distributions.}\label{fig4}
\end{figure}

We also study the effect of the mixture of carbon isotopes $^{12}$C and $^{13}$C on the thermal conductivity of GNRs. Here, we demonstrate the results in the case of armchair GNRs (qualitatively similar results are obtained for zigzag GNRs). The
concentration of $^{13}$C is $N_{13}/(N_{12}+N_{13})$, where $N_{12/13}$ is the number of $^{12/13}$C atoms. The thermal
conductivity is seen to be reduced by introducing $^{13}$C, and the thermal conductivity of pure $^{13}$C GNRs is lower than that of pure
$^{12}$C GNRs because $^{13}$C atoms have larger mass and thus give lower phonon frequency.\cite{Ashcroft}
The inset of Fig.~\ref{fig3}(a) shows a typical random isotope distribution. Another isotope
distribution pattern we study is the isotopic supperlattice structure shown in the inset of Fig.~\ref{fig3}(b). Here, the whole GNR is composed of four slices with equal length and the same isotope composition. Within each slice (such as the dashed box in the inset of Fig.~\ref{fig3}(b)), the number ($L$) of vertical $^{13}$C atomic chains with zigzag
shape (e.g., $L=4$ for the inset of Fig.~\ref{fig3}(b)) can vary from 0 to 7 (see details in Ref.~\onlinecite{supp}). $L=0$ ($L=7$) corresponds to the pure $^{12}$C ($^{13}$C) GNR. The temperature dependent thermal conductivity in Fig.~\ref{fig3} shows
that the isotope effect becomes more evident at higher temperatures. We show the thermal conductivity as a function of
the concentration of $^{13}$C calculated at the temperature of 500 K in
Fig.~\ref{fig4}. In the case of the random
distributions, the calculated thermal conductivity is an average of 10 different 
distributions with the same isotope concentration. For random isotope distributions, the
isotope concentration dependent thermal conductivity (dashed line in Fig.~\ref{fig4}) shows a pan shape and is relatively flat in the
concentration range of $\sim$20-90\%. The thermal conductivity is reduced by $\sim$10\% around the isotope concentration of $\sim$50\%. The error bars are determined from
the deviations of the thermal
conductivities for the 10 different distributions
from their average value. In contrast, the conical shape of
the solid line in Fig.~\ref{fig4} shows much stronger dependence of the
thermal conductivity on the isotope concentration for the supperlattice
structures, with $\sim$30\% reduction of the thermal conductivity at
$\sim$50\% of the isotope concentration. We have also obtained similar results\cite{unpub} using velocity exchange MD\cite{Muller97} in the LAMMPS package.\cite{Plimpton95}

It has been previously demonstrated
that the thermal conductivity depends on the edge chirality\cite{Hu09,Jiang09,Xu09}
in the absence of H-passivation, i.e., the thermal conductivity of the zigzag
GNR is larger than that of the armchair GNR. However, as shown in Fig.~\ref{fig2}, their thermal
conductivities become close to each other within the MD error
bars after the H-passivation, suggesting that phonon scattering with
the hydrogenated edges dominates over the contribution from the chirality effect. We have
suggested that the smaller thermal conductivity of armchair GNRs is due
to the stronger phonon scattering at the armchair edges.\cite{Hu09} It is
interesting to note that the H-passivated zigzag GNR in Fig.~\ref{fig1}(b) resembles the armchair GNRs at the edges. We suggest that the thermal
transport in small GNRs (several nanometers in size in our study) is
strongly affected by the edge configuration.

Recently, it has
been experimentally demonstrated that different carbon isotopes can be controllably introduced in graphene,
such as $^{13}$C, in the chemical vapor deposition growth of graphene on
metals. Both random and segregation (by domains of different isotopes) distributions have been observed.\cite{Li09b} This opens possibilities of engineering the thermal properties of
graphene by isotope distributions. The isotope effect on the thermal transport has been studied in
several nanomaterials, such as carbon nanotubes,\cite{Stoltz09} boron nitride nanotubes,\cite{Chang06b, Stewart09} and silicon nanowires.\cite{Yang08} The pan shape of the dashed line in Fig.~\ref{fig4} is consistent with the the reduction of the thermal conductivity in the ``alloy limit."\cite{Kim06} A similar pan shape is found in GNRs\cite{Jiang10} and SiGe nanowires\cite{Chen09} by tuning the composition. 
By keeping the isotope
concentration a constant of $\sim$50\%, it has been shown\cite{Ouyang09} that the thermal conductivity as a function of the slice
length (which is kept constant in our simulations) gives similar
conical shaped dependence as we see in Fig.~\ref{fig4} for the supperlattice structures.

In conclusion, the classical MD is applied to calculate
the thermal conductivities of rectangular GNRs. We
find that the edge H-passivation can reduce the thermal conductivity significantly. We also show that the thermal conductivity depends on the concentrations of isotopic atoms and their
distribution patterns. The isotopic supperlattice distributions can reduce the
thermal conductivity much more than random distributions.
These findings can be useful in controlling heat
transfer in nanoscale using GNR-based thermal devices.

This work is partially supported by the Semiconductor Research
Corporation (SRC) - Nanoelectronics Research Initiative (NRI) via
Midwest Institute for Nanoelectronics Discovery (MIND) and by Cooling Technologies Research Center (CTRC) at Purdue University.

%

\section*{Supplementary Material}
Supperlattice structures of different $^{13}$C concentrations are shown in the following plots. The larger (smaller) symbols 
denote the $^{13}$C ($^{12}$C) atoms. The number $L$, denoting the number of columns of zigzag $^{13}$C 
atomic chains in each subplot, varies from 0 to 6 for these plots.
\begin{figure}[h]
\includegraphics[width=0.4\textwidth]{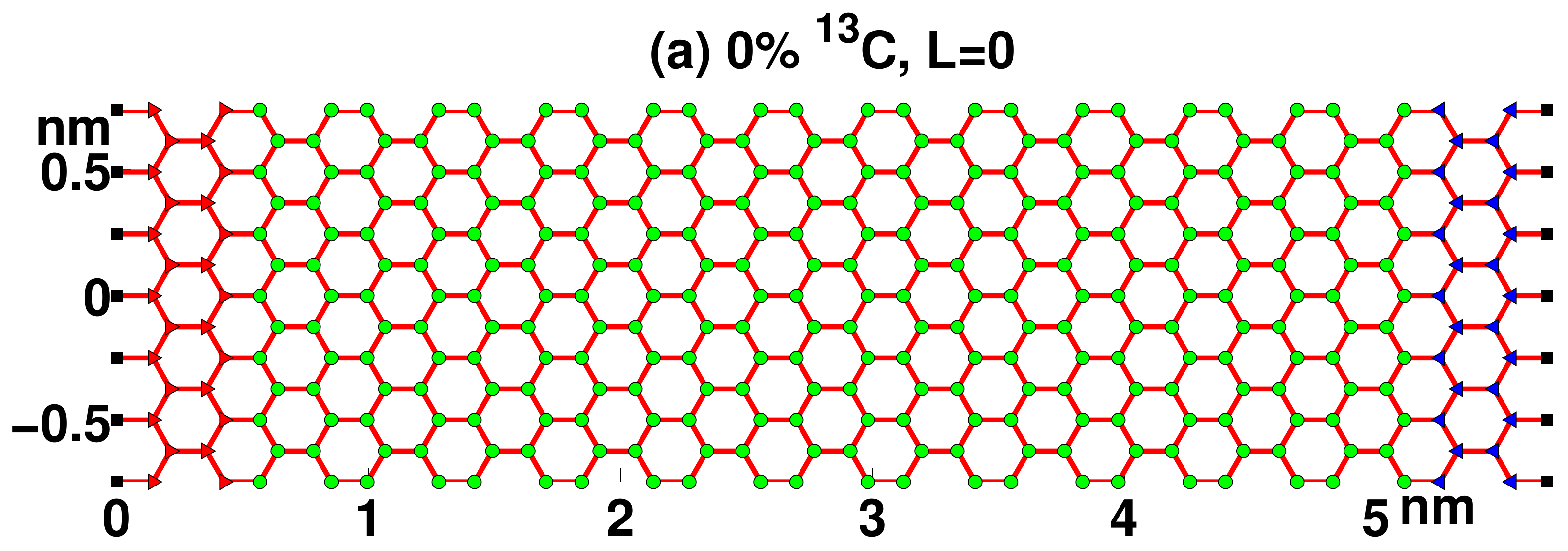}
\includegraphics[width=0.4\textwidth]{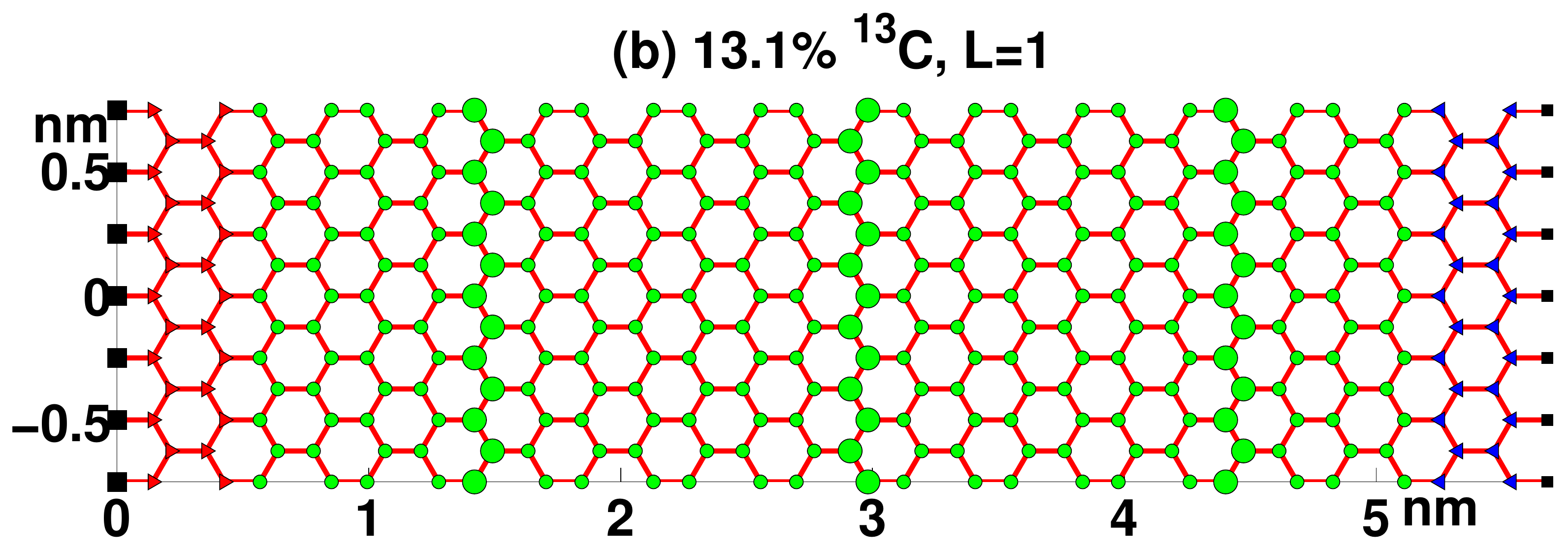}
\includegraphics[width=0.4\textwidth]{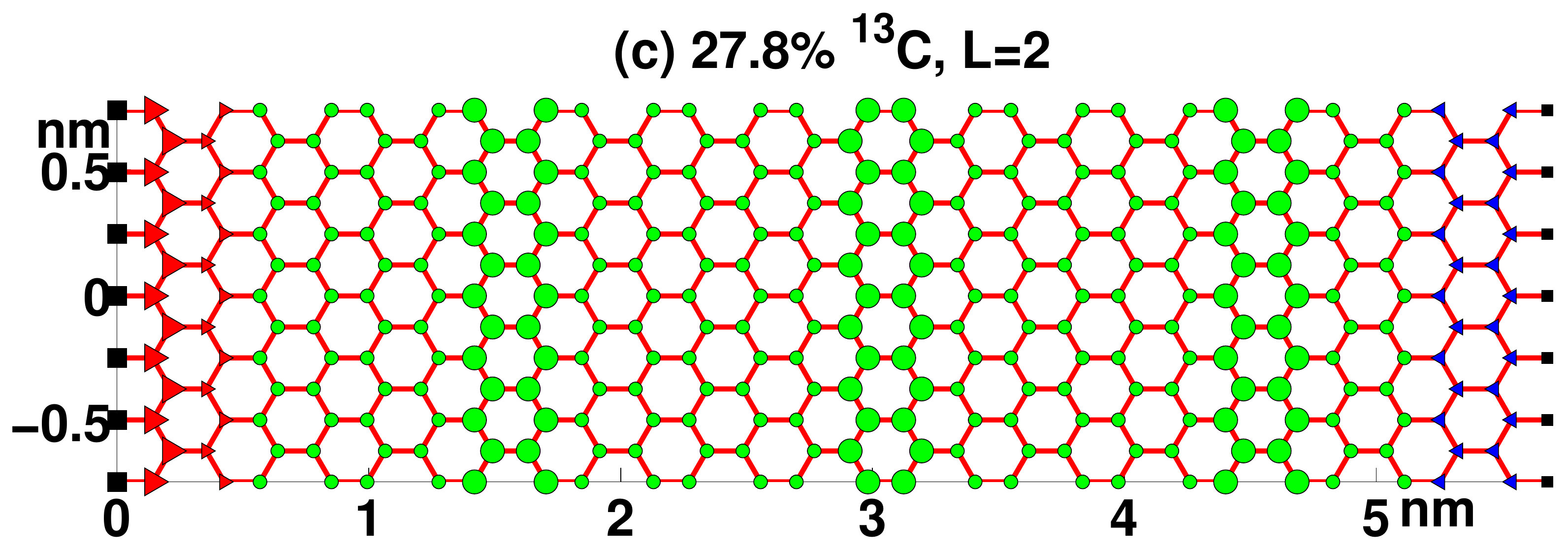}
\end{figure}
\begin{figure}[h]
\includegraphics[width=0.4\textwidth]{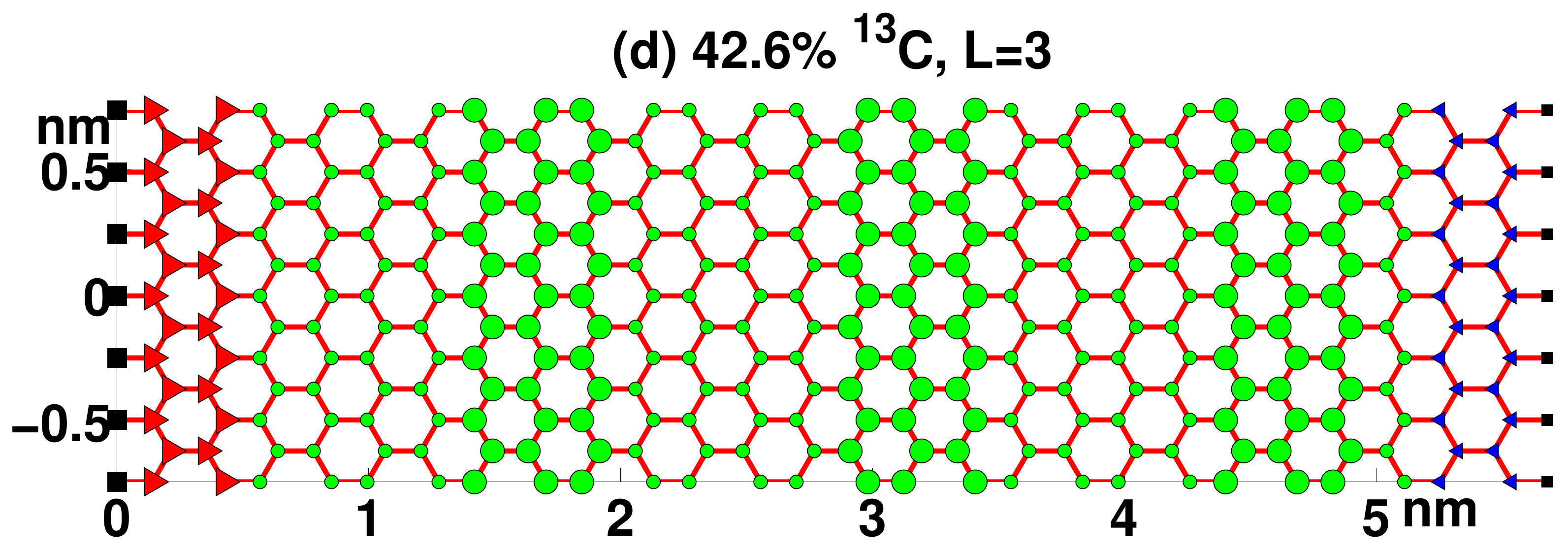}
\includegraphics[width=0.4\textwidth]{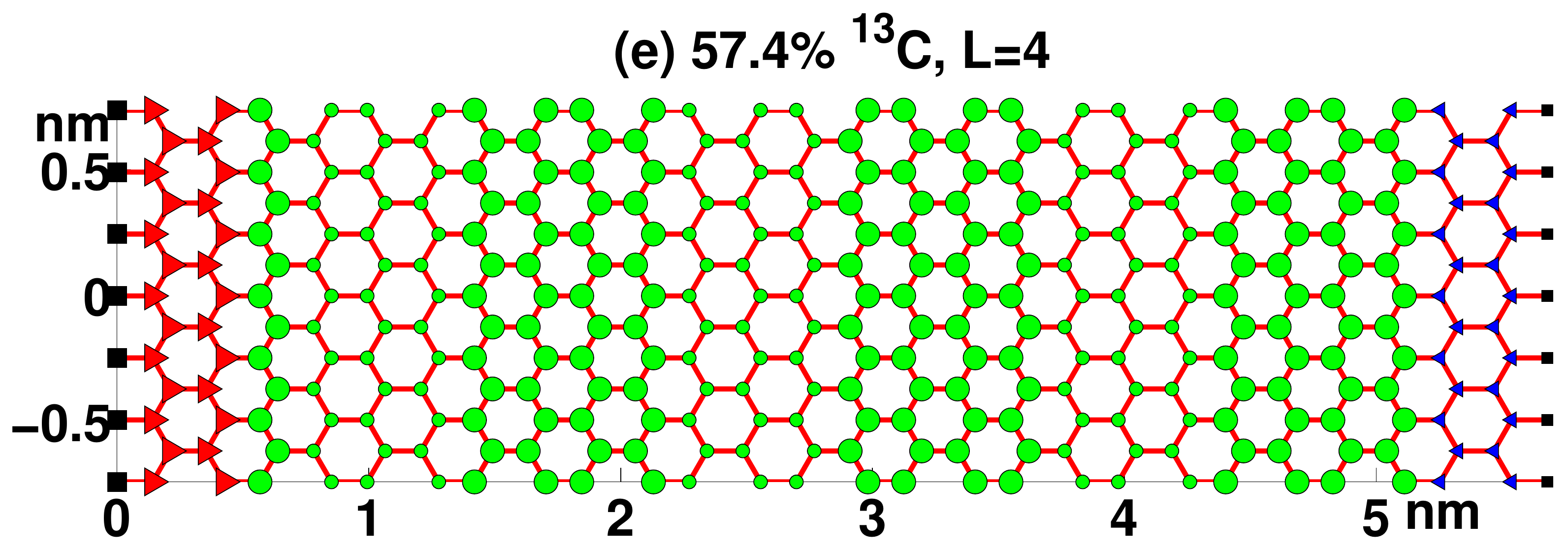}
\includegraphics[width=0.4\textwidth]{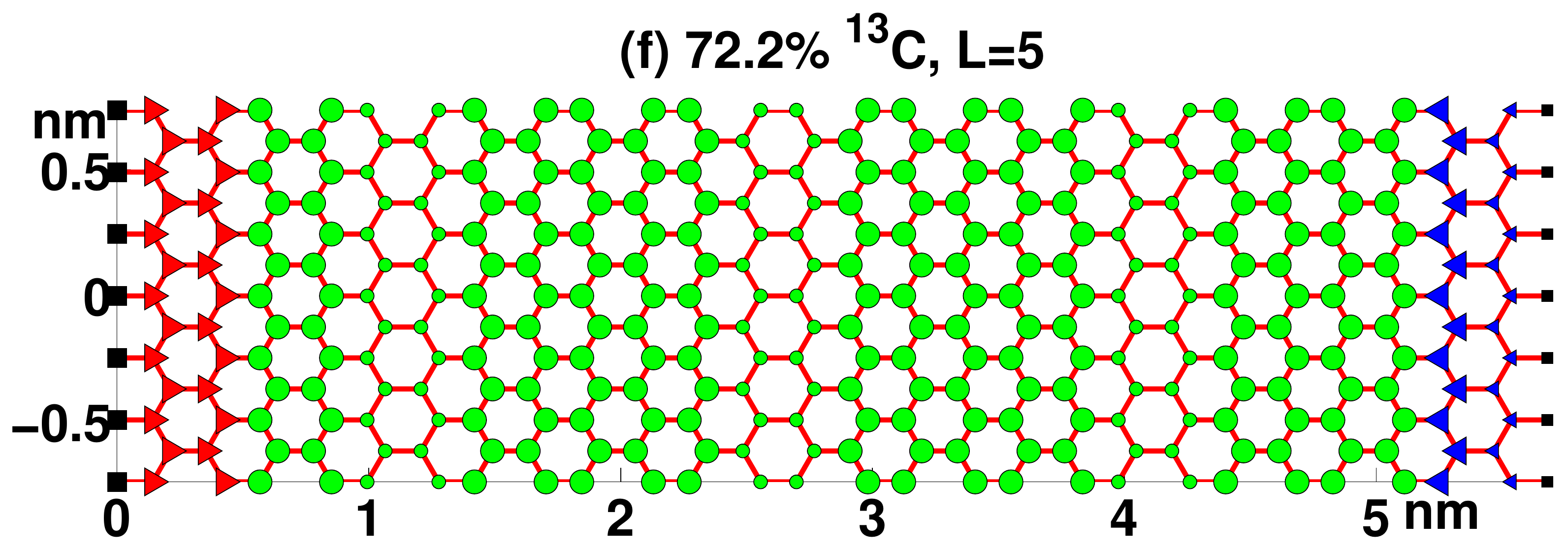}
\includegraphics[width=0.4\textwidth]{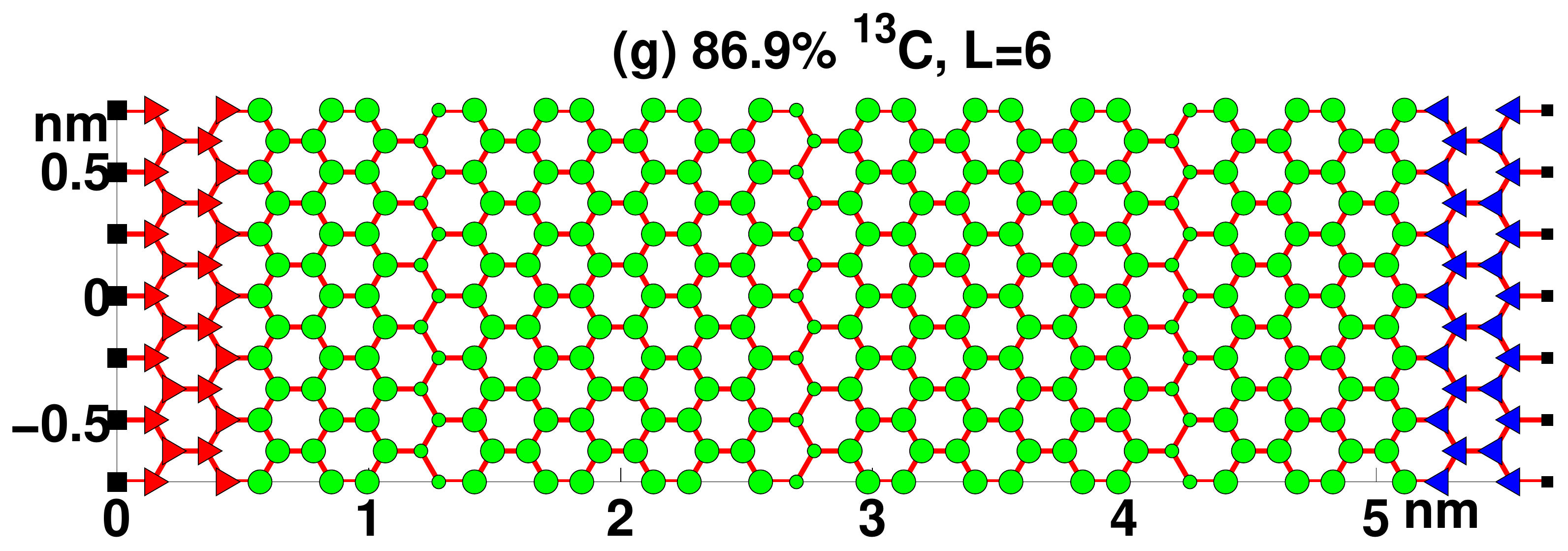}
\end{figure}
\end{document}